\documentclass[pdflatex,sn-nature]{sn-jnl}

\usepackage{graphicx}
\usepackage{amsmath,amssymb,amsfonts}
\usepackage{booktabs}
\usepackage{xcolor}
\usepackage{CJKutf8}
\hypersetup{hidelinks}
\setcitestyle{super,sort&compress}
\makeatletter
\renewcommand{\abstract}[1]{\gdef\@abstract{\@parboxrestore\normalfont\normalsize\bfseries\noindent#1\par}}
\makeatother

\begin{document}

\title{Amplified transport variability in granular slope flows near the critical angle}

\author*[1]{\fnm{Fumiaki} \sur{Nakai}}\email{fumiaki.nakai@ess.sci.osaka-u.ac.jp}
\author*[1]{\fnm{Yuto} \sur{Sasaki}}\email{sasaki.geoscience@gmail.com}

\affil[1]{\orgdiv{Department of Earth and Space Science},
\orgname{The University of Osaka},
\orgaddress{\street{1-1 Machikaneyama}, \city{Toyonaka},
\postcode{560-0043}, \state{Osaka}, \country{Japan}}}

\abstract{
Mechanical disturbances such as seismic shaking can set a granular
slope in motion. Understanding the resulting transport before arrest
is important for predicting topographic change, assessing landslide
hazards and controlling powder flows in industrial processes.
Experiments show that smaller disturbances
can trigger avalanches and larger areas are mobilized as slopes approach
their stability limit \citep{DaerrDouady1999,DeboeufEtAl2003}.
Yet how much transport varies between events under the same macroscopic
conditions and forcing remains unclear. Here we use numerical simulations
to show that transport variability is amplified near the critical angle.
We apply identical basal pulses to static granular layers with the same
macroscopic conditions but different initial particle configurations.
Although the mean transport increases approximately in inverse proportion
to the angular distance from the critical angle, the standard deviation
increases even faster and can exceed the mean. Thus, identical forcing can
produce widely different transport amounts. Deterministic constitutive
laws that assign a unique response to given macroscopic conditions
cannot capture this variability. We introduce fluctuations into the
energy balance to develop a stochastic model that captures the main
features of the broadening transport distribution. These results show
that assessing transport near the critical angle requires considering
its distribution, not just its mean, and motivate constitutive
descriptions that incorporate fluctuations.
}

\maketitle

Understanding how granular slopes move and come to rest after a mechanical
disturbance is important for predicting changes in topography caused by
seismic shaking and for landslide hazard assessment
\citep{AndreottiEtAl2013,Keefer2002,FanEtAl2019,JerolmackDaniels2019,RoeringEtAl2001}.
Granular layers can remain static below a critical slope angle $\theta_c$,
known as the angle of repose, but a disturbance can induce transient flow~\citep{AndreottiEtAl2013,DeboeufEtAl2003,PouliquenForterre2002}.
This response depends strongly on slope angle
\citep{Pouliquen1999,SilbertEtAl2003,KamrinHenann2015}.
Experiments on granular slopes have shown that smaller disturbances can
trigger avalanches as the stability limit is approached, and that
avalanche propagation varies with slope angle and layer thickness
\citep{DaerrDouady1999}.
As the angle of repose is approached, relaxation after an avalanche
takes longer, and a localized impact mobilizes grains over a larger area
\citep{DeboeufEtAl2003}.
These results show that slopes near their stability limit are more
easily set in motion, with the resulting movement lasting longer
and extending over larger areas.

Granular flows can exhibit substantially different dynamics
from trial to trial as they come to rest, even under the same
macroscopic conditions
\citep{PicaCiamarraEtAl2011,PastoreEtAl2011}.
Numerical simulations of granular materials under constant shear stress
and pressure have shown that arrest times are broadly distributed just
below the critical stress, with their mean and variance diverging as
the stress approaches the critical shear stress from below
\citep{SrivastavaEtAl2019}.
How such variability between trials is reflected in the amount of
transport generated by mechanically triggered granular slope flows
remains unclear.
In particular, when macroscopically identical slopes are subjected to
the same disturbance, it remains unresolved under what conditions
the variability in transport before arrest increases and how its
distribution can be described theoretically.

Here we use discrete element simulations to study the statistics of
transport from flow onset to arrest in granular slopes subjected to
identical basal pulses. We keep macroscopic conditions, including slope
angle and system geometry, fixed across trials and vary only the
microscopic particle configurations.
As the slope angle $\theta$ approaches the critical angle $\theta_c$,
the mean transport increases approximately in inverse proportion to
the angular distance $\theta_c-\theta$.
The standard deviation, however, grows with a larger power-law exponent
and can exceed the mean.
Thus, transport increases near the critical angle, but so does its
variability relative to the mean.
Deterministic descriptions that assign a unique response to given
macroscopic conditions cannot capture this variation between trials.
To describe this variability, we develop a model that incorporates
stochastic fluctuations into the energy balance.
The model broadly captures the increase in the mean and standard
deviation, as well as the probability distribution of transport
at different slope angles.
These results show that assessing granular transport in this regime
requires considering its distribution, not just its mean, and that
a theoretical description of this distribution requires a stochastic
framework.

\section*{Numerical set-up and transport measure}

We used discrete element simulations to examine how an initially static
granular layer responds to a single basal pulse on a rough slope
(Fig.~\ref{fig:q-time}a).
Particles were modelled as spheres, with their
centres constrained to move in the plane defined by the downslope
and slope-normal directions.
The layer contained 2,273 mobile particles
with diameters uniformly distributed between 1.0 and 1.9 mm and rested
on a base made of particles. The downslope system length was
$L_x=100\,\mathrm{mm}$, or $L_x/d=100$, where
$d=1.0\,\mathrm{mm}$ is the reference particle diameter.
Periodic boundary conditions were applied in the
downslope direction.
To induce flow, we moved the base downslope and back to its original
position over one sinusoidal cycle. We then held the base stationary and tracked
the subsequent particle motion. We varied the slope angle from
$10.0^\circ$ to $16.3^\circ$ and performed 128 trials at each angle.
Across trials at a given angle, we kept the macroscopic conditions
and forcing identical and varied only the initial particle configuration.
Further simulation details are provided in Methods.

We quantified transport using the particle displacements weighted by
their masses and normalized by the downslope system length:
\begin{equation}
 Q(t)=\frac{1}{L_x}\sum_i m_i\Delta x_i(t),
 \label{eq:cumulative-transport}
\end{equation}
where $m_i$ is the mass of particle $i$ and
$\Delta x_i(t)=x_i(t)-x_i(0)$ is its downslope displacement
from the start of forcing, with $x_i(t)$ tracked continuously
across periodic boundaries. The sum includes all mobile particles.
We refer to $Q(t)$ as the cumulative transport.
Once the layer has come to rest, we write the final
transport simply as $Q$.
We normalize $Q$ by $\rho d^3$, where
$\rho=1000\,\mathrm{kg\,m^{-3}}$ is the particle mass density
and $d$ is the reference particle diameter defined above.

\section*{Transport following a single disturbance}

\begin{figure}[tbp]
\centering
\includegraphics[width=\textwidth]{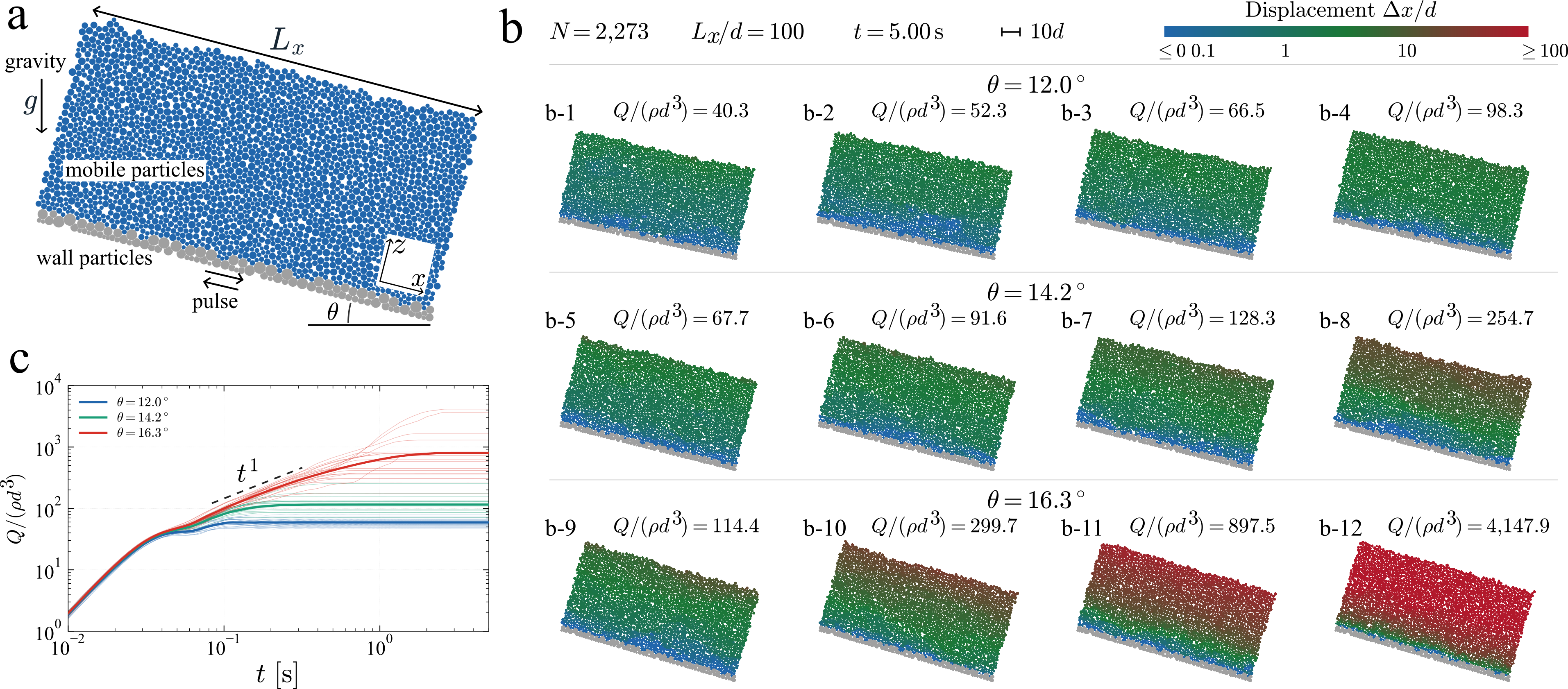}
\caption{\textbf{Granular transport following a single basal pulse.}
\textbf{a}, Initial configuration of the simulated layer
($L_x/d=100$, $\theta=14.2^\circ$). Mobile particles (blue) rest on a rough
base of wall particles (grey), which is moved downslope and back once to
apply the pulse. Gravity $g$ is vertical, and $x$ and $z$ denote the
downslope and slope-normal directions. The downslope boundaries are periodic.
\textbf{b}, Final configurations 5 s after the onset of the pulse at
$\theta=12.0^\circ$ (b-1--b-4), $14.2^\circ$ (b-5--b-8) and $16.3^\circ$
(b-9--b-12), for a layer of $N=2273$ mobile particles. At each angle,
the four panels show the trials ranked 1, 32, 96 and 128 by $Q$ among 128
trials. Colours indicate the downslope displacement $\Delta x/d$ from the
onset of the pulse on a logarithmic scale, and the value of $Q/(\rho d^3)$
is given above each panel. Both the transport and its spread among trials
increase with slope angle (see also Supplementary Movie 1).
\textbf{c}, Normalized cumulative transport
$Q(t)/(\rho d^3)$ against time $t$ from the onset of the pulse
on logarithmic axes, for $L_x/d=100$ at $12.0^\circ$ (blue), $14.2^\circ$
(green) and $16.3^\circ$ (red). Thin lines show 20 trials with different
initial configurations, and thick lines show the mean over 128 trials. The
dashed line is a guide proportional to $t^1$. Each trial approaches
a plateau, and both the mean and the spread increase with slope angle.}
\label{fig:q-time}
\end{figure}

A single basal pulse induces downslope motion.
Figure~\ref{fig:q-time}b shows the resulting particle displacements
at $t=5\,\mathrm{s}$.
We compare three
representative slope angles, $12.0^\circ$, $14.2^\circ$, and
$16.3^\circ$, for $L_x/d=100$.
At each angle, the four panels in Fig.~\ref{fig:q-time}b show
trials with relatively small to large final transport.
Figure~\ref{fig:q-time}c shows how transport develops over time:
thin lines represent 20 trials with different initial particle
configurations, and thick lines show the mean over 128 trials.
In each trial shown, $Q(t)$ approaches a plateau, indicating that
net downslope transport has largely ceased. The plateau values
increase on average with slope angle.

The spread in final transport across trials increases with slope angle.
The common colour scale in Fig.~\ref{fig:q-time}b allows particle
displacements to be compared across trials and angles. By the time
shown, net downslope transport has effectively ceased in all these
trials (see also Supplementary Movie~1).
At $12.0^\circ$, the selected
trials show relatively small differences in displacement
(Fig.~\ref{fig:q-time}b-1--b-4). At $16.3^\circ$, the displacement
magnitudes differ substantially between trials
(Fig.~\ref{fig:q-time}b-9--b-12), even though the slope angle and
basal pulse are identical. These differences are also evident in
Fig.~\ref{fig:q-time}c: the plateau values of $Q$ are closely
grouped at $12.0^\circ$ but span a wider range at $16.3^\circ$.
Thus, increasing the slope angle increases both the mean final
transport and its variability across initial particle configurations.
We next examine the statistical properties of the final transport.

\section*{Broadening of transport distributions}

\begin{figure}[tbp]
\centering
\includegraphics[width=\textwidth]{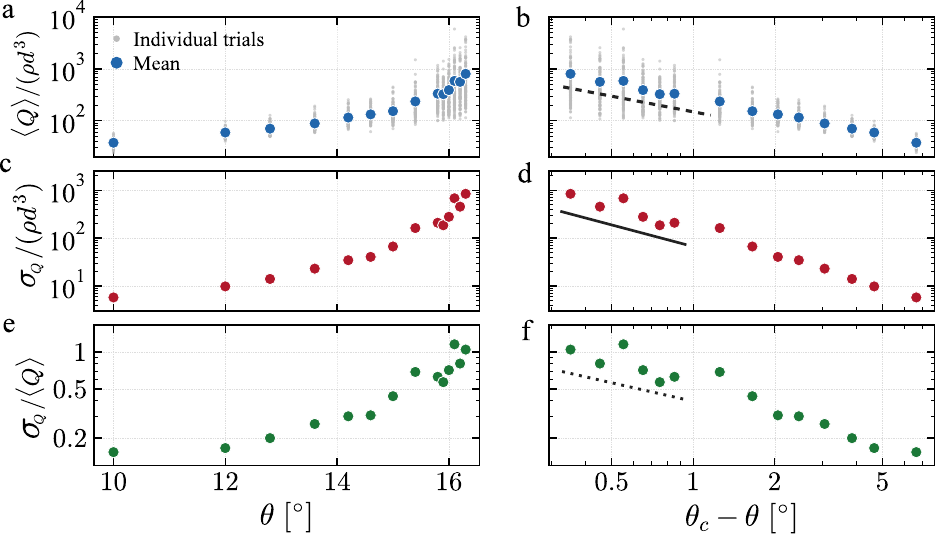}
\caption{\textbf{Angle dependence of the mean and variability of transport.}
Cumulative transport $Q$ was evaluated 5 s after the onset of forcing
for $L_x/d=100$, with 128 trials at each of 14 slope angles between
$10.0^\circ$ and $16.3^\circ$.
Panels \textbf{a,b} show the normalized mean
$\langle Q\rangle/(\rho d^3)$ (blue), \textbf{c,d} the normalized
standard deviation $\sigma_Q/(\rho d^3)$ (red), and \textbf{e,f}
the ratio $\sigma_Q/\langle Q\rangle$ (green).
Grey points in \textbf{a,b} show individual trials.
The left column plots these quantities against the slope angle
$\theta$, and the right column against the distance
$\theta_c-\theta$ from the critical angle
$\theta_c\simeq16.65^\circ$.
Both the mean and the standard deviation increase as the critical
angle is approached. Their ratio also increases, indicating that
the standard deviation grows faster than the mean.
The black segments are guides showing the power-law scalings
predicted by our model:
$(\theta_c-\theta)^{-1}$ in \textbf{b} (dashed),
$(\theta_c-\theta)^{-1.5}$ in \textbf{d} (solid), and
$(\theta_c-\theta)^{-0.5}$ in \textbf{f} (dotted).}
\label{fig:q-statistics}
\end{figure}

The mean cumulative transport increases as the slope angle approaches the
critical angle. Figure~\ref{fig:q-statistics}a,b shows the normalized
mean transport $\langle Q\rangle/(\rho d^3)$ for $L_x/d=100$,
evaluated 5 s after the onset of forcing. We performed 128 trials
with different initial particle configurations at each of 14 slope
angles between $10.0^\circ$ and $16.3^\circ$.
The mean cumulative transport rises increasingly steeply as the slope approaches
the critical angle $\theta_c\simeq16.65^\circ$.
This trend agrees qualitatively with experiments showing that,
as the angle of repose is approached, localized impacts mobilize
larger areas and relaxation after an avalanche takes longer
\citep{DeboeufEtAl2003}.

The spread around the mean also increases towards the critical angle.
The grey points in Fig.~\ref{fig:q-statistics}a,b show that transport
amounts under the same conditions span a wider range at larger slope
angles. The standard deviation increases
(Fig.~\ref{fig:q-statistics}c,d), and its ratio to the mean,
$\sigma_Q/\langle Q\rangle$, approaches and can exceed unity near
the critical angle (Fig.~\ref{fig:q-statistics}e,f).
Variations between trials are then comparable to or larger than
the mean transport. The mean alone therefore cannot adequately
describe individual events in this regime; the distribution of
transport amounts must also be considered.
The variability also depends on system size: the standard deviation
decreases approximately as $(L_x/d)^{-1/2}$ with increasing
system length $L_x/d$.
Nevertheless, its increase towards the critical angle is observed
for all three system sizes studied
(Appendix Fig.~\ref{fig:q-statistics-size}).

The mean, standard deviation and their ratio increase approximately
as power laws of the angular distance $\theta_c-\theta$.
The guides in Fig.~\ref{fig:q-statistics}b,d,f have exponents
$-1$, $-3/2$ and $-1/2$, respectively, as derived below
(Eq.~\eqref{eq:transport-moments}), and broadly agree with
the trends in the data.

\begin{figure}[tbp]
\centering
\includegraphics[width=\textwidth]{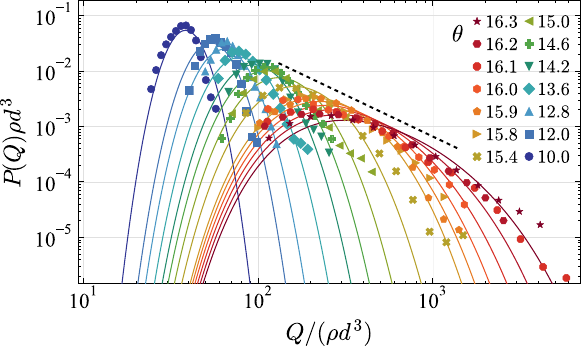}
\caption{\textbf{Distributions of cumulative transport following a single pulse.}
Cumulative transport $Q$ was evaluated 5 s after the onset of forcing
for $L_x/d=100$. The dimensionless probability density $P(Q)\rho d^3$
is plotted against $Q/(\rho d^3)$.
Colours and symbols indicate 14 slope angles from $10.0^\circ$ to
$16.3^\circ$. Symbols show probability densities estimated from
128 trials at each angle using Gaussian kernel density estimation.
As the slope angle increases, the mean transport increases and the
tail extends towards larger transport amounts.
Solid lines show the theoretical model
(Eq.~\eqref{eq:transport-distribution}), which broadly captures these trends.
A joint fit to the raw samples at all angles, with the critical angle
fixed at $\theta_c\simeq16.65^\circ$, gives the common parameters
in Eq.~\eqref{eq:transport-distribution}, expressed in dimensionless form as
$K_0/(\rho g d^3)=4.899$ and $D/(\rho g^2d^4)=1.066$.
The dashed line shows the asymptotic scaling $Q^{-1.5}$
predicted by the model in the critical angle limit.}
\label{fig:q-distribution}
\end{figure}

To examine the distribution of transport across trials,
Fig.~\ref{fig:q-distribution} shows the dimensionless probability
density $P(Q)\rho d^3$. For $L_x/d=100$, we evaluated cumulative transport $Q$
5 s after the onset of forcing, using 128 trials at each slope angle
from $10.0^\circ$ to $16.3^\circ$. Probability densities were estimated
using Gaussian kernels in the logarithm of normalized transport
(Methods).

As the slope angle increases, the mean transport increases and the
tail extends towards larger transport amounts. At low angles,
transport values are concentrated within a narrow range. At high
angles, some trials produce transport amounts well above the mean,
even under identical macroscopic conditions and forcing.
This broadening towards larger transport amounts corresponds to
the increase in the ratio of the standard deviation to the mean
shown in Fig.~\ref{fig:q-statistics}.
Solid lines show the distribution derived from the theoretical model
developed below (Eq.~\eqref{eq:transport-distribution}).
With a single set of parameters fitted jointly across all angles,
the model broadly captures the increase in mean transport and the
broadening of the tail.
The dashed line shows the asymptotic scaling $Q^{-1.5}$ predicted
by the model in the critical limit.

\section*{A theoretical description of transport statistics}

We develop a theoretical model to explain the increases in mean transport
and variability between trials observed in the simulations as the
critical angle is approached.
We first adopt a constant-friction approximation to $\mu$--$I$ rheology
\citep{Pouliquen1999,GDRMiDi2004,daCruzEtAl2005,JopEtAl2006,ForterrePouliquen2008}
and calculate transport before arrest from a deterministic energy
balance between gravitational work and frictional dissipation.
This description gives a unique transport
amount for given conditions and therefore cannot capture the spread
of the distribution. We then introduce fluctuations phenomenologically into the energy
balance and treat arrest as a first-passage problem, in which the
energy first reaches zero. This approach yields a transport
distribution that broadly captures the trends observed in the simulations.
The assumptions and derivations are given in
Appendix~\ref{app:transport-theory}.

The deterministic energy balance accounts for the increase in mean
transport near the critical angle.
We represent the disturbance by an initial kinetic energy, without
resolving the finite duration of the pulse, and approximate the bulk
friction coefficient in $\mu$--$I$ rheology by a constant $\mu_0$.
Let $K$ denote the downslope kinetic energy per unit system length
and $Q$ the transport, with initial conditions $K=K_0>0$ and $Q=0$.
The equation of motion (Eq.~\eqref{eq:deceleration}) gives
$dK=-a\,dQ$, where $a=g(\mu_0\cos\theta-\sin\theta)$ is the net
energy loss per unit transport and is positive for
$\theta<\theta_c=\arctan\mu_0$.
The layer stops when $K$ reaches zero, giving $Q=K_0/a$
(Eq.~\eqref{eq:appendix-deterministic-transport}).
For fixed $K_0$, transport therefore increases approximately
in inverse proportion to $\theta_c-\theta$ near the critical
angle, as observed for the mean transport in the simulations.

To describe the variability between trials, we introduce fluctuations
phenomenologically into the energy balance.
We assume that the variance of the accumulated random
work is proportional to transport and that correlations between
work contributions decay rapidly with spatial separation and
as transport progresses. We approximate the sum of many such
contributions as Gaussian and neglect its short memory along
the transport coordinate, giving
\begin{equation}
 dK=-a\,dQ+b\,dW_Q,
 \qquad b^2=\frac{2D}{L_x},
 \label{eq:stochastic-energy-balance}
\end{equation}
where $D$ is the noise strength and $W_Q$ is a standard Wiener
process with transport $Q$ as its independent variable.
Its increments are
Gaussian with zero mean and variance $dQ$.
The factor $L_x^{-1}$ arises from spatial averaging.
Starting from $K=K_0>0$ at $Q=0$, transport before arrest is determined
by a first-passage problem \citep{vanKampen2007,Risken1996,LimaEtAl2000}:
the energy reaches zero for the first time.
Solving the Fokker--Planck equation corresponding to
Eq.~\eqref{eq:stochastic-energy-balance}, with an absorbing boundary
at $K=0$, gives the following inverse-Gaussian distribution of
transport before arrest:
\begin{equation}
 P(Q)
 =\frac{K_0}{\sqrt{2\pi b^2Q^3}}
 \exp\!\left[-\frac{(K_0-aQ)^2}{2b^2Q}\right],
 \qquad Q>0.
 \label{eq:transport-distribution}
\end{equation}

We compare the predicted distribution with the total transport
measured in the simulations, including transport during the pulse.
For this comparison, $K_0$ is treated as an effective initial
energy rather than the kinetic energy measured at the end of
the pulse.
With $\theta_c=16.65^\circ$ fixed, we fitted
Eq.~\eqref{eq:transport-distribution} jointly to the raw transport
values across all 14 angles by maximum likelihood, using common
values of $K_0$ and $D$.
Without adjusting these parameters separately for each angle,
the model broadly captures both the shift towards larger transport
amounts and the broadening of the tail as the slope angle increases
(Fig.~\ref{fig:q-distribution}).

Equation~\eqref{eq:transport-distribution} gives the main properties
of the mean transport, its variability and the distribution tail.
The mean and variance follow as the first and second cumulants,
respectively (Appendix~\ref{app:transport-theory}):
\begin{equation}
 \langle Q\rangle=\frac{K_0}{a},
 \qquad
 \sigma_{Q}^{2}
 =\frac{b^2K_0}{a^3}
 =\frac{2DK_0}{L_xa^3}.
 \label{eq:transport-moments}
\end{equation}
The mean equals the transport predicted by the deterministic energy
balance (Eq.~\eqref{eq:appendix-deterministic-transport}),
while fluctuations give rise to a nonzero variance.
For fixed $K_0$, $D$ and $L_x$, we have
$a\propto\theta_c-\theta$ near the critical angle.
The mean therefore diverges as $(\theta_c-\theta)^{-1}$,
whereas the standard deviation diverges more rapidly,
as $(\theta_c-\theta)^{-3/2}$.
Their ratio consequently diverges as $(\theta_c-\theta)^{-1/2}$.
As the critical angle is approached, gravitational
work nearly balances mean frictional dissipation, making transport
increasingly sensitive to fluctuations. This broadly explains why
the spread across trials becomes comparable to, and can exceed,
the mean transport in the simulations
(Fig.~\ref{fig:q-statistics}).
At fixed slope angle $\theta$, effective initial energy $K_0$
and noise strength $D$, the standard deviation decreases as
$L_x^{-1/2}$, in broad agreement with the observed system-size
dependence (Fig.~\ref{fig:q-statistics-size}).
The distribution tail has an exponential cutoff below the critical
angle and follows $Q^{-3/2}$ asymptotically at $\theta=\theta_c$.
The simulations at high angles suggest a similar trend, although
a clear power-law regime is not resolved
(Fig.~\ref{fig:q-distribution}).

\section*{Implications for transport predictability}

Near the critical angle, the mean alone cannot adequately characterize
transport on granular slopes. Slopes with identical macroscopic
conditions but different microscopic particle configurations produce
widely different amounts of transport under the same disturbance
(Fig.~\ref{fig:q-time}).
As the critical angle is approached, the standard deviation grows
faster than the mean and can exceed it
(Fig.~\ref{fig:q-statistics}).
Thus, the increase in mean transport is accompanied by an increase
in relative variability, with the distribution extending towards
much larger transport amounts (Fig.~\ref{fig:q-distribution}).
Although our simulations concern idealized granular slopes, these
findings suggest that microscopic differences in internal structure
could also contribute to variability in landslide transport, even
when slope angles, material properties and external forcing are
similar. Assessing slope transport near the stability limit therefore
requires considering the distribution of transport amounts, not
just their mean.

Describing transport variability in more general systems calls for
incorporating fluctuations into constitutive laws. Deterministic
descriptions based on $\mu$--$I$ rheology~\citep{Pouliquen1999,JopEtAl2006,ForterrePouliquen2008,KamrinHenann2015,MowlaviKamrin2021,KimKamrin2020} give a unique response
for the same initial and boundary conditions and therefore cannot
capture variability arising from microscopic configurations that
are not resolved in the macroscopic fields. Fluctuations in sheared
granular materials have been reported
\citep{MillerEtAl1996,SasakiKatsuragi2025,SantosEtAl2022}, 
and their role in macroscopic flow has been explored
through models that link stochastic rearrangements at the mesoscale
to mean flow fields
\citep{PouliquenForterre2009,KamrinBazant2007}.
However, these models have not linked such rearrangements to
variability between trials in transport.
Here we introduced fluctuations phenomenologically into a spatially
coarse-grained energy balance
(Eq.~\eqref{eq:stochastic-energy-balance}, with details in
Appendix~\ref{app:transport-theory}).
The resulting transport distribution and its mean and variance
(Eqs.~\eqref{eq:transport-distribution}
and~\eqref{eq:transport-moments}) broadly describe the variability
observed in the simulations. Our model relies on the energy
balance of the whole layer in the present slope geometry, and extending
it to more general systems is not straightforward. Incorporating
fluctuations into local constitutive laws could allow both the mean
response and its variability to be described across different
geometries and boundary conditions. 
Developing stochastic constitutive laws that capture variability
between trials is an important direction for granular rheology.

\bibliography{main_ref}

\section*{Methods}

We used discrete element method (DEM) simulations in LAMMPS~\citep{ThompsonEtAl2022} to
examine how transport before arrest varies between granular slopes
subjected to identical disturbances under the same macroscopic
conditions. At each slope angle, we performed 128 trials, keeping
macroscopic conditions, including particle number, fixed and varying
only the microscopic initial particle configurations. We applied an
identical basal pulse to each initially static layer and tracked its
motion from flow onset to arrest. We repeated these simulations at
14 slope angles between $10.0^\circ$ and $16.3^\circ$ to examine
transport statistics and their dependence on slope angle. The main
text presents results for $L_x/d=100$, where $L_x$ is the downslope
system length and $d=1.0\,\mathrm{mm}$ is the reference particle
diameter. Additional results for $L_x/d=30$ and $300$ are presented
in the Appendix to assess system-size dependence. The simulation
code and input files will be deposited on Zenodo and made publicly available.

The system consisted of a layer of mobile particles resting on a
rough base made of particles (Fig.~\ref{fig:q-time}a).
Particles were treated as spheres, with their centres constrained
to move in the plane defined by the downslope and slope-normal
directions. We applied periodic boundary conditions in the downslope
direction and fixed the relative positions of the base particles.
For $L_x/d=100$, the layer contained 2,273 mobile particles with
diameters drawn from a uniform distribution between 1.0 and 1.9 mm
and a material density of $1000\,\mathrm{kg\,m^{-3}}$.
We calculated particle translation and rotation under gravity and
contact forces. Here, $g$ denotes the gravitational
acceleration. Contact forces were calculated using the
Hertz--Mindlin--Tsuji model.
We set Young's modulus to $1.0\,\mathrm{MPa}$,
Poisson's ratio to 0.48, the interparticle friction coefficient to 0.6 and the
restitution coefficient parameter to 0.40. A low Young's modulus
was used to reduce the computational cost. The time step was
$10^{-6}\,\mathrm{s}$.

In each trial, we prepared and relaxed the granular layer before
applying a basal pulse. We first deposited particles onto the base
and smoothly reduced the slope angle from $30^\circ$ to the target
angle over 7 s by changing the direction of gravity. We then allowed
the layer to relax at the target angle for a further 13 s.
To induce flow, we moved the base downslope and back to its original
position in a single pulse. The basal displacement was
$u_{\mathrm b}(t)=A[1-\cos(2\pi t/T)]/2$ for $0\leq t\leq T$,
where $t$ is the time since the onset of forcing.
The maximum displacement was $A=1.0\,\mathrm{mm}$ and the pulse
duration was $T=0.0635\,\mathrm{s}$, corresponding to a frequency
of approximately 15.8 Hz and a maximum acceleration of $0.5g$.
After the pulse, we held the base stationary and tracked particle
displacements for the remainder of the 5 s observation period.
We used the transport measured 5 s after the onset of forcing
for the statistical analysis.

The probability densities in Fig.~\ref{fig:q-distribution} were
estimated from 128 trials at each angle using Gaussian kernels in
$\ln[Q/(\rho d^3)]$. The bandwidth was $h=1.06\sigma_{\ln Q}n^{-1/5}$,
where $\sigma_{\ln Q}$ is the sample standard deviation of
$\ln[Q/(\rho d^3)]$ and $n=128$.
We divided the density estimated in logarithmic space by $Q$
to obtain $P(Q)$ and displayed it over the observed range.

\subsection*{Data availability}

All data required to reproduce the results of this study, including the
source data underlying the figures, will be deposited on Zenodo and made
publicly available. The DOI will be added in a future version of this manuscript.

\subsection*{Code availability}

All simulation and analysis code, input files and documentation required to
reproduce the results of this study will be deposited on Zenodo and made
publicly available. The DOI will be added in a future version of this manuscript.

\section*{Acknowledgements}

Computations were performed using the facilities of the Supercomputer
Center, the Institute for Solid State Physics, the University of Tokyo
(ISSPkyodo-SC-2026-Ba-0021).

\section*{Funding}

This work was supported by JSPS KAKENHI (grant nos.\ JP24KJ0156 and
JP25K17359).

\section*{Competing interests}

The authors declare no competing interests.

\section*{Additional information}

Correspondence and requests for materials should be addressed to F.N. or Y.S.

\clearpage
\begin{appendices}
\renewcommand{\theHfigure}{appendix.\arabic{figure}}
\renewcommand{\theHequation}{appendix.\thesection.\arabic{equation}}
\section{System-size dependence of transport statistics}
\label{app:q-size-dependence}

\begin{figure}[htbp]
\centering
\includegraphics[width=\textwidth]{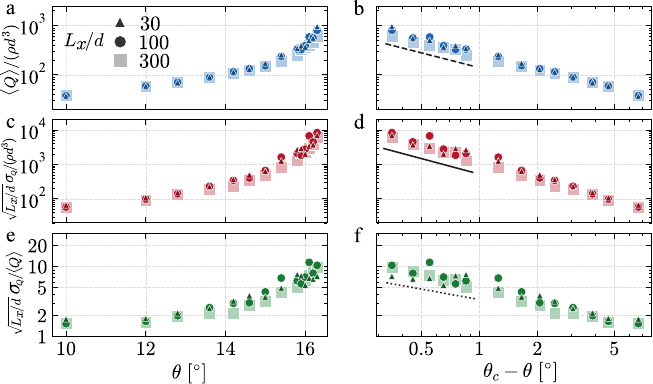}
\caption{\textbf{System-size dependence of transport statistics.}
Results are shown for $L_x/d=30$ (triangles), 100 (circles) and
300 (squares). Transport $Q$ was evaluated 5 s after the onset
of forcing, using 128 trials at each of 14 slope angles between
$10.0^\circ$ and $16.3^\circ$ for each system size.
\textbf{a,b}, Normalized mean transport
$\langle Q\rangle/(\rho d^3)$ (blue).
\textbf{c,d}, Scaled standard deviation
$\sqrt{L_x/d}\,\sigma_Q/(\rho d^3)$ (red).
\textbf{e,f}, Scaled ratio of the standard deviation to the mean,
$\sqrt{L_x/d}\,\sigma_Q/\langle Q\rangle$ (green).
The left column shows the dependence on slope angle $\theta$,
and the right column shows the same data against the angular
distance $\theta_c-\theta$, with $\theta_c\simeq16.65^\circ$.
The mean transport varies relatively little with system size,
whereas the standard deviation decreases in larger systems.
Multiplying the standard deviation by $\sqrt{L_x/d}$ approximately
collapses the data for the three sizes, indicating
$\sigma_Q\propto(L_x/d)^{-1/2}$. The similarly scaled ratio to the
mean also approximately collapses.
For all three sizes, the mean, standard deviation and their ratio
increase towards the critical angle.
The black segments are guides showing the power-law scalings
predicted by our model:
$(\theta_c-\theta)^{-1}$ in \textbf{b} (dashed),
$(\theta_c-\theta)^{-3/2}$ in \textbf{d} (solid), and
$(\theta_c-\theta)^{-1/2}$ in \textbf{f} (dotted).}
\label{fig:q-statistics-size}
\end{figure}

\section{Energy balance and first-passage description of transport}
\label{app:transport-theory}

\subsection{Deterministic energy balance}

We first calculate the total transport before arrest of a granular layer
set in motion by a mechanical disturbance. We approximate the bulk
friction coefficient in the $\mu$--$I$ constitutive law by a constant
and derive the transport from the energy balance between gravitational
work and frictional dissipation. We represent the disturbance by
an initial kinetic energy and follow the motion of the layer over
the fixed base until arrest, without resolving the finite duration
of the disturbance.

We derive the deceleration of the layer by substituting the
constant-friction constitutive relation into the momentum balance.
Let $x$ and $z$ denote the downslope and slope-normal coordinates,
respectively. We assume parallel flow
$\boldsymbol{v}=v(z,t)\boldsymbol{e}_x$ with constant layer thickness
and bulk density $\rho_b$, neglecting slope-normal motion and
dilation. In sheared regions with $\partial_zv\geq0$,
$\mu$--$I$ rheology relates the shear stress $\tau$ to the
slope-normal pressure $p_n$ through $\tau=\mu(I)p_n$, where
$I=d\,\partial_zv/\sqrt{p_n/\rho}$ is the inertial number,
$d$ is the reference particle diameter and $\rho$ is the
particle material density. We neglect the dependence of friction
on the inertial number and approximate $\mu(I)\simeq\mu_0$.
Because the flow is uniform in $x$, the advective acceleration
vanishes, and the momentum balance becomes
\begin{equation}
 \rho_b\partial_t v=\rho_b g\sin\theta+\partial_z\tau,
 \qquad
 \partial_zp_n=-\rho_b g\cos\theta.
 \label{eq:appendix-momentum}
\end{equation}
Substituting $\tau=\mu_0p_n$ gives
\begin{equation}
 \partial_t v=-a,
 \qquad
 a=g(\mu_0\cos\theta-\sin\theta).
 \label{eq:deceleration}
\end{equation}
For $\theta<\theta_c=\arctan\mu_0$, we have $a>0$, so moving
regions decelerate at a constant rate. We assume that each region
remains at rest once its velocity reaches zero. In these static
regions, the stress satisfies force balance rather than the
constitutive relation used during flow.

We use this deceleration law to derive an energy balance linking
the kinetic energy of the layer to transport. Let $K(t)$ denote
the downslope translational kinetic energy per unit system length
and $Q(t)$ the cumulative transport. For a layer of length $L_x$,
thickness $H$ and total mass $M$, we use the reference particle
diameter $d$ as the effective out-of-plane width, giving
$\rho_b=M/(L_xHd)$. The kinetic energy and transport rate are then
\begin{equation}
 \begin{aligned}
 K(t)
 &=\frac{\rho_b d}{2L_x}
   \int_0^{L_x}\!\int_0^H v^2\,dz\,dx
 =\frac{M}{2L_xH}\int_0^H v^2\,dz,\\
 \dot Q(t)
 &=\frac{\rho_b d}{L_x}
   \int_0^{L_x}\!\int_0^H v\,dz\,dx
 =\frac{M}{L_xH}\int_0^H v\,dz.
 \end{aligned}
 \label{eq:appendix-energy-definitions}
\end{equation}
The final expressions use the assumed uniformity of the flow in $x$.
We neglect particle rotational kinetic energy.
Multiplying Eq.~\eqref{eq:deceleration} by $Mv/(L_xH)$ and
integrating over the moving regions gives
\begin{equation}
 \dot K
 =g\sin\theta\,\dot Q-\mu_0g\cos\theta\,\dot Q
 =-a\dot Q.
 \label{eq:appendix-energy-balance}
\end{equation}
The first term on the right is the rate of work done by gravity,
and the second is the rate of energy loss through friction.
Static regions contribute to neither term because their velocity
is zero.

The initial conditions and the arrest condition determine the total
transport before the layer comes to rest. Integrating
Eq.~\eqref{eq:appendix-energy-balance} with $K(0)=K_0>0$ and
$Q(0)=0$ gives $K=K_0-aQ$. At arrest, $K=0$, giving the total transport
\begin{equation}
 Q=\frac{K_0}{a}
 =\frac{K_0\cos\theta_c}{g\sin(\theta_c-\theta)}
 \simeq\frac{K_0\cos\theta_c}{g(\theta_c-\theta)}.
 \label{eq:appendix-deterministic-transport}
\end{equation}
The final approximation holds as $\theta$ approaches $\theta_c$
from below, with the angular distance expressed in radians.
For fixed initial energy $K_0$, the total transport therefore
increases approximately in inverse proportion to the angular
distance from the critical angle, consistent with the mean
transport in Fig.~\ref{fig:q-statistics}b.

\subsection{Stochastic energy balance}

To describe transport variability between trials, we extend the
deterministic energy balance $\dot K=-a\dot Q$
(Eq.~\eqref{eq:appendix-energy-balance}) to include fluctuating work
associated with contact-network rearrangements.
Multiplying this balance by $dt$ and using $dK=\dot K\,dt$
and $dQ=\dot Q\,dt$ gives $dK=-a\,dQ$.
We then add a random work increment $d\eta$ phenomenologically
to obtain the coarse-grained balance for the whole layer:
\begin{equation}
 dK=-a\,dQ+d\eta,
 \qquad
 a=g(\mu_0\cos\theta-\sin\theta),
 \label{eq:appendix-fluctuating-balance}
\end{equation}
Here, $K$ is the kinetic energy per unit downslope length,
$Q$ is the cumulative transport, and $d\eta$ is the random work
increment per unit length over the same time interval $dt$.
The deterministic term accounts for gravitational work and mean
frictional dissipation, while $d\eta$ represents the fluctuating work.

We model the fluctuating work as the sum of contributions from
contact-network rearrangements. We assume that the number of
rearrangements is proportional to the total mass-weighted particle
displacement, $L_x\,dQ$. We also assume that the work associated
with each rearrangement has finite variance and that correlations
between the work contributions decay sufficiently rapidly with
spatial separation and as transport progresses. The variance of
the summed work is then proportional to $L_x\,dQ$.
Since $d\eta$ is this work normalized by $L_x$, its variance is divided
by $L_x^2$ and is therefore proportional to $dQ/L_x$.
Including the mean resistance in $a$, we model the mean and variance
of each work increment, conditional on the preceding state, as
\begin{equation}
 \langle d\eta\rangle=0,
 \qquad
 \langle(d\eta)^2\rangle=\frac{2D}{L_x}\,dQ,
 \label{eq:appendix-work-variance}
\end{equation}
where $D$ is a coefficient that combines the effects of the variance
of the work contributions from contact-network rearrangements
and the correlations between them.
At fixed layer thickness $H$, we assume that $D$ is independent
of slope angle and system length.
We approximate the summed work from many rearrangements as a
Gaussian process and its short correlations along the transport
coordinate by white noise. The work increment can then be written as
\begin{equation}
 d\eta=\sqrt{\frac{2D}{L_x}}\,dW_Q,
 \label{eq:appendix-work-wiener}
\end{equation}
where $W_Q$ is a standard Wiener process indexed by transport $Q$.
We assume that $Q$ increases monotonically during flow.
Substituting this expression into
Eq.~\eqref{eq:appendix-fluctuating-balance} gives
\begin{equation}
 dK=-a\,dQ+\sqrt{\frac{2D}{L_x}}\,dW_Q.
 \label{eq:appendix-transport-noise}
\end{equation}
With the initial conditions $K(0)=K_0>0$ and $Q(0)=0$,
we identify arrest with the first time the energy reaches zero.
The transport before arrest can then be determined from this
first-passage problem.

\subsection{Transport distribution and moments}

We derive the distribution of transport before arrest from the
stochastic energy balance
(Eq.~\eqref{eq:appendix-transport-noise}).
Let $p(K,Q)\,dK$ be the probability that the layer is still moving
and has energy between $K$ and $K+dK$ after transport $Q$.
With $b^2=2D/L_x$, this energy balance gives the
Fokker--Planck equation \citep{Risken1996,vanKampen2007}
\begin{equation}
 \partial_Qp=a\,\partial_Kp+\frac{b^2}{2}\partial_K^2p,
 \qquad
 p(K,0)=\delta(K-K_0),
 \qquad
 p(0,Q)=0.
 \label{eq:appendix-fokker-planck}
\end{equation}
The first term on the right describes the mean decrease in energy.
The second describes how fluctuations spread the energy distribution.
All trials start with energy $K_0>0$, and the absorbing boundary
at $K=0$ means that motion stops when the energy reaches zero.
Requiring $p$ and $\partial_Kp$ to vanish as $K\to\infty$ gives
\begin{equation}
 \begin{aligned}
 p(K,Q)=\frac{1}{\sqrt{2\pi b^2Q}}
 \Bigg\{&
 \exp\!\left[-\frac{(K-K_0+aQ)^2}{2b^2Q}\right]\\
 &-e^{2aK_0/b^2}
 \exp\!\left[-\frac{(K+K_0+aQ)^2}{2b^2Q}\right]
 \Bigg\}.
 \end{aligned}
 \label{eq:appendix-absorbing-solution}
\end{equation}
The integral $\int_0^\infty p(K,Q)\,dK$ is the probability that
the layer is still moving after transport $Q$.
Its decrease as $Q$ increases gives the probability density
of transport before arrest:
\begin{equation}
 \begin{aligned}
 P(Q)
 &=-\frac{d}{dQ}\int_0^\infty p(K,Q)\,dK
 =\frac{b^2}{2}\partial_Kp(0,Q)\\
 &=\frac{K_0}{\sqrt{2\pi b^2Q^3}}
 \exp\!\left[-\frac{(K_0-aQ)^2}{2b^2Q}\right].
 \end{aligned}
 \label{eq:appendix-first-passage-flux}
\end{equation}
This is an inverse-Gaussian distribution. For $a>0$, it integrates
to one, meaning that the layer eventually stops with probability one.

The mean and variance of transport follow from the Laplace
transform of the distribution,
\begin{equation}
 F(s)=\left\langle e^{-sQ}\right\rangle
 =\exp\!\left[
 \frac{K_0}{b^2}
 \left(a-\sqrt{a^2+2b^2s}\right)
 \right].
 \label{eq:appendix-laplace}
\end{equation}
The mean is $-F'(0)$ and the variance is
$F''(0)-[F'(0)]^2$, giving
\begin{equation}
 \langle Q\rangle=\frac{K_0}{a},
 \qquad
 \sigma_Q^2=\frac{2DK_0}{L_xa^3},
 \qquad
 \frac{\sigma_Q}{\langle Q\rangle}
 =\sqrt{\frac{2D}{L_xK_0a}}.
 \label{eq:appendix-transport-moments}
\end{equation}
For fixed $K_0$ and $D$, the relation
$a\propto\theta_c-\theta$ near the critical angle implies that
the mean, standard deviation and their ratio grow as
$(\theta_c-\theta)^{-1}$, $(\theta_c-\theta)^{-3/2}$ and
$(\theta_c-\theta)^{-1/2}$, respectively.
These scalings broadly agree with the trends in
Fig.~\ref{fig:q-statistics}b,d,f.
At fixed angle, the standard deviation and its ratio to the mean
both decrease as $L_x^{-1/2}$, broadly consistent with the system-size
dependence in Fig.~\ref{fig:q-statistics-size}.
Equation~\eqref{eq:appendix-first-passage-flux} also predicts
an asymptotic $P(Q)\propto Q^{-3/2}$ tail
at the critical angle, where $a=0$.
The simulations at high slope angles show a broadly similar trend
(Fig.~\ref{fig:q-distribution}).

\end{appendices}

\end{document}